# Simulation Modeling

by Florian Hartig

## Introduction

With the rise of computers, simulation models have emerged beside the more traditional statistical and mathematical models as a third pillar for ecological analysis. Broadly speaking, a simulation model is an algorithm, typically implemented as a computer program, which propagates the states of a system forward. Unlike in a mathematical model, however, this propagation does not employ the methods of calculus but rather a set of rules or formulae that directly prescribe the next state. Such an algorithmic model specification is particularly suited for describing systems that are difficult to capture or analyze with differential equations such as: (a) systems that are highly nonlinear or chaotic; (b) discrete systems, for example networks or groups of distinct individuals; (c) systems that are stochastic; and (d) systems that are too complex to be successfully treated with classical calculus. As these situations are frequently encountered in ecology, simulation models are now widely applied across the discipline. They have been instrumental in developing new insights into classical questions of species' coexistence, community assembly, population dynamics, biogeography, and many more. The methods for this relatively young field are still being actively developed, and practical work with simulation models requires ecologists to learn new skills such as coding, sensitivity analysis, calibration, validation, and forecasting uncertainties. Moreover, scientific inquiry with complex systems has led to subtle changes to the philosophical and epistemological views regarding simplicity, reductionism, and the relationship between prediction and understanding.

## General Overviews

Three short overview articles that jointly paint a good picture of the field are Jackson, et al. 2000 on ecological modeling, Pascual 2005 on computational approaches in ecology, and Black and McKane 2012 on stochastic simulations in ecology. Huston, et al. 1988 is another short piece that not only provides a good overview about questions and aims of simulation models in ecology but also an explanation of why these aims cannot be realized with simpler mathematical models (about mathematical models, see also the separate *Oxford Bibliographies* article Mathematical Ecology). Two useful textbooks on ecological modeling are Jørgensen and Bendoricchio 2001 and Grimm and Railsback 2005. Jørgensen and Bendoricchio 2001 provides an introduction to ecological modeling, leaning toward system analysis and system models, a topic also reviewed in the separate *Oxford Bibliographies* article Systems Ecology. Grimm and Railsback 2005 focuses on individual-based models in ecology. Two further technical references are Zeigler, et al. 2000, a comprehensive general introduction to simulation modeling, and Wilkinson 2011, an excellent technical reference on stochastic simulations. Finally, a note: simulations are also frequently employed in statistical methods, for example in statistical null models (e.g., Gotelli 2000). Such approaches, however, which only resample or simulate data without describing an explicit ecological process, are not covered in this article.

**Black, A. J., and A. J. McKane. 2012. Stochastic formulation of ecological models and their applications.** *Trends in Ecology & Evolution* 27.6: 337–345.

A recent review on stochastic simulation models in ecology, with a focus on individual-based simulations.

**Gotelli, N. J. 2000. Null model analysis of species co-occurrence patterns.** *Ecology* 81.9: 2606–2621.

A classic reference on statistical null models that employ simulation or resampling algorithms to compare an observed pattern to a null expectation.

**Grimm, V., and S. F. Railsback. 2005.** *Individual-based modeling and ecology*. Princeton, NJ: Princeton Univ. Press.

This standard textbook provides an excellent introduction to the field of individual-based models in ecology.

**Huston, M., D. DeAngelis, and W. Post. 1988. New computer models unify ecological theory.** *BioScience* 38.10: 682–691.

A short piece that not only provides a good overview about questions and aims of complex simulation models in ecology but also an explanation of why these aims cannot be realized with simpler mathematical models.

**Jackson, L. J., A. S. Trebitz, and K. L. Cottingham. 2000. An introduction to the practice of ecological modeling.** *BioScience* 50.8: 694–706.

A short and gentle introduction into the topics of ecological modeling (mathematical and simulation models).

**Jørgensen, S. E., and G. Bendoricchio. 2001.** *Fundamentals of ecological modelling*. 4th ed. Amsterdam, The Netherlands: Elsevier.

A classic textbook on ecological modeling, with a focus on system analysis and system models.

**Pascual, M. 2005. Computational ecology: From the complex to the simple and back.** *PLoS Computational Biology* 1.2: e18.

A short introduction to the field of computation ecology, with tree examples that show how simulation models can help to understand complex adaptive ecological systems.

**Wilkinson, D. J. 2011.** *Stochastic modelling for systems biology*. Boca Raton, Florida: CRC Press.

Although this book is written primarily for system biologists, it will also be useful to ecologists, through its thorough introduction to the field of stochastic simulations techniques.

**Zeigler, B. P., H. Praehofer, and T. G. Kim. 2000.** *Theory of modeling and simulation: Integrating discrete event and continuous complex dynamic systems*. Cambridge, Massachusetts, United States: Academic Press.

A classic textbook on simulation models in science in general.

## Journals

Studies using simulation models are published in virtually all ecology journals, as well as in other general-topic journals that publish ecological studies. As a tendency, however, simulation models will more likely be found in journals with a focus on advancing ecological theory and synthesis, in particular *Ecology Letters*, *American Naturalist*, *Ecography*, *Oikos*, and *Theoretical Ecology*. Also, there are a number of journals that concentrate on methodological aspects and modeling. Of those, *Ecological Modelling* arguably has the strongest focus on traditional, larger simulation models. This journal is therefore a typical place to present new model descriptions or applications. Other methodological journals of note are *Environmental Modelling and Software*, and *Methods in Ecology and Evolution*. Moreover, at the intersection of ecology and evolution, journals such as *Systematic Biology* or *Evolution* frequently publish simulation-based studies.

*The American Naturalist*.

Looks back on a proud tradition as one of the leading ecology journals since its inception in 1867. The journal focuses on studies that advance general ecological theory.

*Ecography*.

A well-respected journal concentrating on spatial and temporal patterns in ecology from the community to the macroscale. It has a relatively strong focus on synthesis and theory, and it frequently publishes simulation models on those topics.

*Ecology Letters*.

As one of the most respected general ecology journals, *Ecology Letters* publishes concise articles that promise to spark general interest and advance the field of ecology. The journal regularly publishes studies based on simulation models, as long as they are of sufficiently broad interest to ecologists.

*Ecological Modelling*.

The main journal for the presentation of techniques and applications in ecological modeling. For studies on modeling techniques, or on the application of a model in a particular system, *Ecological Modelling* is a natural choice.

*Environmental Modeling and Software*.

Similar to *Ecological Modelling* but with a stronger focus on larger environmental models and environmental informatics.

*Evolution*.

Publishes studies on all aspects of evolution, including simulation models that concentrate on this field.

*Methods in Ecology and Evolution*.

A relatively young journal, this has swiftly become the prime outlet for publishing methodological advances in ecology. Although the journal concentrates mainly on statistical and empirical methods, it also publishes simulation frameworks or methods for the analysis of simulation models and sometimes applications of simulation models.

*Oikos*.

A respected journal that publishes research on all aspects of ecology, with a focus on work aimed at generalization and synthesis.

*Systematic Biology*.

This is the journal of the Society of Systematic Biologists and a leading journal on all aspects of systematics, including phylogeny, trait evolution, and biogeography. It frequently publishes simulation models on these topics.

*Theoretical Ecology*.

A relatively new journal that has placed itself at the intersection between the more traditional mathematical ecology journals and the modeling journals, such as *Ecological Modelling* and *Environmental Modelling*, and *Software*. *Theoretical Ecology* tends to have more of a strong theoretical character than *The American Naturalist* or *Oikos*.

## Important Concepts and Principles in the Field of Simulation Models

Over the last few centuries, science has relied mainly on mathematics as a tool and reductionism as an epistemological method. It is therefore not surprising that the emergence of computer simulations to study complex systems sparked discussions about the philosophical underpinnings of this new field. The aim of this section is to summarize central technical concepts of the field and how they relate to epistemological thought in science.

## Discrete Entities: Cellular Automaton, Networks, and Individual-based Models

Being able to simulate according to algorithms provides a much more general way to study the consequences of interactions between discrete entities than previously existing methods from statistical mechanics. Examples of simulations that concentrate on discrete entities are cellular automata (CA, see, e.g., Wolfram 2002), network models (see, e.g., Albert and Barabási 2002), and individual-based models (IBMs), also known as agent-based models (ABMs) in the social sciences (see, e.g., Grimm 1999). While Wolfram 2002 may have slightly exaggerated calling this revolution "a new kind of science," the ability to simulate discrete systems has proven immensely useful for ecology. Judson 1994 provides an excellent account of the new possibilities that discrete entity models introduced to ecology at that time. Today, CA, network, and IBM models have become fixed items in an ecologist's toolbox. They are also heavily used in the study of human-environmental interactions (e.g., Bousquet and Page 2004). Yet, modelers also miss the analytical elegance and generality of mathematical models, and some effort has been made to find approximate analytical solutions for discrete simulation models (see, e.g., Murrell, et al. 2004).

**Albert, R., and A. L. Barabási. 2002. Statistical mechanics of complex networks.** *Reviews of Modern Physics* **74.1: 47.**

This often-cited review provides an excellent overview about simulation models and results in the field of complex networks.

**Bousquet, F., and C. Le Page. 2004. Multi-agent simulations and ecosystem management: A review.** *Ecological Modelling* **176.3: 313–332.**

A review on the use of agent-based simulations in ecosystem management.

**Grimm, V. 1999. Ten years of individual-based modelling in ecology: What have we learned and what could we learn in the future?** *Ecological Modelling* **115.2: 129–148.**

This review, although somewhat dated, summarizes motivation and emerging results from individual-based models in ecology.

**Judson, O. P. 1994. The rise of the individual-based model in ecology.** *Trends in Ecology & Evolution* **9.1: 9–14.**

Provides an accessible introduction to the motivations for moving from mathematical models to individual-based simulations.

**Murrell, D. J., U. Dieckmann, and R. Law. 2004. On moment closures for population dynamics in continuous space.** *Journal of Theoretical Biology* **229.3: 421–432.**

A good overview about methods to approximate spatiotemporal dynamics of individual-based models through analytical equations.

**Wolfram, S. 2002.** *A new kind of science*. **Champaign, IL: Wolfram Media.**

Provides an extensive account of the developments and promises in the field of cellular automata.

## Complexity and Emergence

The ability to describe and simulate new structures went hand in hand with a revolution in thinking about natural systems. This revolution was characterized by the realization that simple interaction rules between local entities can lead to complicated dynamics and the emergence of surprising macroscopic patterns—as well as by the idea that complex systems may reorganize themselves to react to internal or external drivers. Consequently, scientists turned away from trying to understand nature by reducing each problem to the smallest possible entities (methodological reductionism), and instead started to study larger systems as a whole, leading to the field of complex systems science. The problem of inferring the underlying local processes from observed large-scale patterns is discussed in the classic and still highly relevant Levin 1992, which phrases this problem of as one of the main challenges for ecological research. The second idea, that systems can reorganize their structure in response to internal or external drivers, is summarized in the idea of "complex adaptive systems" (see Holland 1992). Levin 1998 reviews and refines the concept for ecology. Interesting in this context is also the earlier, highly influential article Bak, et al. 1987, which introduced the concept of "self-organized criticality": this refers to the phenomenon that complex systems can regulate themselves to maintain a state close to a critical phase transition. A special issue in *Science*, with an editorial by Gallagher, et al. 1999, provides an excellent view into the thinking and the state of the field at the turn of the millennium. Further references can be found in the separate *Oxford Bibliographies* article Complexity Theory.

**Bak, P., C. Tang, and K. Wiesenfeld. 1987. Self-organized criticality: An explanation of the 1/f noise.** *Physical Review Letters* **59.4: 381.**

This paper introduced the idea of self-organized criticality (i.e., the idea that complex systems may regulate themselves to maintain a state close to a critical phase transition, resulting in the emergence of self-similar macroscopic patterns).

**Gallagher, R., T. Appenzeller, and D. Normile. 1999. Beyond reductionism.** *Science* **284.5411: 79.**

Introduction to a special issue in *Science* on the promise of complex systems, which provides a good overview about the thinking and expectations regarding this field at the turn of the century.

**Holland, J. H. 1992. Complex adaptive systems.** *Daedalus* **121.1: 17–30.**

Introduces and defines complex adaptive systems as complex systems with an evolving structure that can reorganize their properties and configuration in response to internal and external drivers.

**Levin, S. A. 1992. The problem of pattern and scale in ecology: The Robert H. MacArthur award lecture.** *Ecology* **73.6: 1943–1967.**

A classic article on the emergence of patterns from lower-level processes and the challenge for ecologists to invert this process and infer the underlying processes from the patterns that we observe.

**Levin, S. A. 1998. Ecosystems and the biosphere as complex adaptive systems.** *Ecosystems* **1.5: 431–436.**

In this influential paper, Levin discusses the idea of complex adaptive systems with a focus on ecological questions.

## Stochasticity and Chaos

Another area that profited enormously from the availability of simulation models are stochastic and highly nonlinear or chaotic systems. Both types of systems can, in principle, be modeled with differential equations, but their solution is often excessively complicated. Simulation models are, therefore, the usual choice if stochasticity is involved, especially when models are highly nonlinear. Wilkinson 2009 provides an excellent introduction to stochastic simulation of biological systems. Particular interest in chaotic systems was sparked by the idea that self-organization may favor evolution toward the "edge of chaos" (see Langton 1990) and that chaotic dynamics should therefore be common in nature. However, while nonlinear and stochastic models are ubiquitous in current research, the interest in chaotic dynamics has somewhat subsided; this is arguably because of the difficulty to distinguish chaos from noise in ecological data and because of theoretical doubts about whether nature is likely to develop chaotic dynamics. A good discussion of this topic can be found in Hastings, et al. 1993. Further references can be found in the separate *Oxford Bibliographies* article Biological Chaos and complex dynamics.

**Hastings, A., C. L. Hom, S. Ellner, P. Turchin, and H. C. J. Godfray. 1993. Chaos in ecology: Is mother nature a strange attractor?** *Annual Review of Ecology and Systematics* **24.1: 1–33.**

Provides an excellent discussion of the concepts of and evidence for chaotic dynamics in ecology.

**Langton, C. G. 1990. Computation at the edge of chaos: Phase transitions and emergent computation.** *Physica D: Nonlinear Phenomena* **42.1–3: 12–37.**

Discusses the tendency of complex systems to evolve toward the "edge of chaos": that is, toward critical phase transitions where self-similar or complex patterns emerge.

**Wilkinson, D. J. 2009. Stochastic modelling for quantitative description of heterogeneous biological systems.** *Nature Reviews Genetics* **10.2: 122–133.**

A concise introduction to stochastic models in systems biology.

## Philosophy, Epistemology, and the "Truth"

Finally, a topic that inevitably arises is what makes a good model and the question of whether a model can ever be "true." In an influential article, Rykiel 1996 expresses the view that models are not built to exactly reproduce all aspects of a natural system but rather to be "valid," meaning that they satisfy certain performance criteria; this view is shared by Oreskes, et al. 1994. It seems, however, that simulation modelers have become bolder over the years. Today, simulation models are frequently used to test or compare ecological hypotheses, which implicitly assumes that they are structurally realistic (for a discussion of structural realism in science, see Ladyman 2007). An interesting paper in this context is Evans, et al. 2013.

**Evans, M. R., V. Grimm, K. Johst, et al. 2013. Do simple models lead to generality in ecology?** *Trends in Ecology & Evolution* **28.10: 578–583.**

A discussion paper about the right level of model complexity. The article argues that ecological systems are inherently complex and that models therefore require a certain minimum level of complexity to be general.

**Ladyman, J. 2007. Structural realism. In** *Stanford Encyclopedia of Philosophy*. **Edited by Edward N. Zalta.**

A review about the concept of structural realism in the philosophy of science, which posits that the structure of the best scientific theories will resemble the true nature of the physical world.

**Oreskes, N., K. Shrader-Frechette, and K. Belitz. 1994. Verification, validation, and confirmation of numerical models in the earth sciences.** *Science* **263.5147: 641–646.**

An influential discussion on what it means to verify and validate a model that applies a rather pragmatic approach to model validation, in the sense that a model is valid if it performs as desired.

**Rykiel, E. J. 1996. Testing ecological models: The meaning of validation.** *Ecological Modelling* **90.3: 229–244.**

A classical paper arguing that models are built for a purpose, not as a representation of "the truth." Hence, a model is valid if it meets specified performance criteria.

## Applications of Simulation Models in the Various Subdisciplines of Ecology

The aim of the following section is to provide a cross-section of model applications in the various subdisciplines of ecology. This collection is not comprehensive but rather has the purpose to show the breadth of applications for simulation models across all fields of ecology.

### Population and Metapopulation Models

Of all areas of ecology, population dynamics is possibly most strongly associated with classical mathematical modeling and theoretical ecology. Indeed, mathematical models are still widely used in the study of population dynamics (see also the separate *Oxford Bibliographies* article Methods in Population Dynamics). Nevertheless, there is a range of applications where simulations have advantages over analytical approaches, in particular for populations that exhibit strong stochasticity or nonlinear dynamics; structured populations, whose individuals have additional properties such as age or different pheno/genotypes; or spatially structured populations or metapopulations. In principle, many of these problems can be solved analytically, but once more processes are included, dynamics are often easier explored by simulation. To study internally structured populations, matrix or integral projection models are frequently employed. Caswell 2001 is an excellent introduction to the former and Ellner and Rees 2006 to the latter. Hanski 1999 provides an introduction to spatially structured populations. An interesting example is Levin, et al. 1984, which uses a simulation model to explore the consequence of different dispersal strategies in patchy environments. Fahrig 2001 highlights the importance of using stochastic simulations for estimating population survival probabilities in conservation ecology. Blasius, et al. 1999 is an example of an ecological simulation model that combines space with nonlinear dynamics. Further population modeling examples can be found in the separate *Oxford Bibliographies* articles Biological Chaos and Complex Dynamics, Population Viability Analysis, Dynamics of Age- and Stage-Structured Populations and Communities and Metapopulations and Spatial Population Processes.

**Blasius, B., A. Huppert, and L. Stone. 1999. Complex dynamics and phase synchronization in spatially extended ecological systems.** *Nature* **399.6734: 354–359.**

This highly cited study examines the origin of persistent, spatially synchronized population cycles with a spatial simulation model.

**Caswell, H. 2001.** *Matrix population models***. 2d ed. Sunderland, MA: Sinauer Associates.**

A standard introduction to matrix models, which describe an internal population structure through discrete population stages (e.g., age classes). To model the population dynamics, matrix models then employ a transition matrix that describes the transition between the stages.

**Ellner, S. P., and M. Rees. 2006. Integral projection models for species with complex demography.** *American Naturalist* **167.3: 410–428.**

An introduction to integral projection models (IPMs). IPMs are essentially matrix models, except that the internal population structure is described by a continuous rather than a discrete variable.

**Fahrig, L. 2001. How much habitat is enough?** *Biological Conservation* **100.1: 65–74.**

A classical study using simulations to determine extinction risks due to demographic stochasticity in spatially structured habitats.

**Hanski, I. 1999.** *Metapopulation ecology***. Oxford: Oxford Univ. Press.**

An introduction to metapopulation ecology, which is the study of population dynamics in spatially structured (patchy) landscapes.

**Levin, S. A., D. Cohen, and A. Hastings. 1984. Dispersal strategies in patchy environments.** *Theoretical Population Biology* **26.2: 165–191.**

A simulation study exploring the effect of different dispersal strategies in patchy environments.

## Community Models, Including Food Web, Vegetation, and Ecosystem Models

A second field where simulation models are frequently used are community models, starting from simple coexistence or predator-prey models over more complex community or feed web models toward global dynamic vegetation and ecosystem models. Reviews on simpler community and coexistence models can be found in the separate *Oxford Bibliographies* articles Competition and Coexistence in Animal Communities. Seidl, et al. 2011 provides a good overview of forest simulators, with a focus on the implementations of disturbances. Higgins and Scheiter 2012 explores vegetation shifts with a complex vegetation models where plant traits adapt dynamically. Ciais, et al. 2005 is a highly cited paper that highlights the importance of dynamic vegetation models to study community-responses to climate and weather events. Harfoot, et al. 2014 is an attempt at creating a global ecosystem model that encompasses all tropic levels.

**Ciais, P., M. Reichstein, and N. Viovy, et al. 2005. Europe-wide reduction in primary productivity caused by the heat and drought in 2003.** *Nature* **437.7058: 529–533.**

An influential study that shows the power of dynamic vegetation models to describe the reaction of plant communities to environmental triggers.

**Harfoot, M. B., T. Newbold, D. P. Tittensor, et al. 2014. Emergent global patterns of ecosystem structure and function from a mechanistic general ecosystem model.** *PLoS Biolology* **12.4: e1001841.**

The description of a global ecosystem model that attempts to describe the interactions between all trophic levels.

**Higgins, S. I., and S. Scheiter. 2012. Atmospheric CO2 forces abrupt vegetation shifts locally, but not globally.** *Nature* **488.7410: 209–212.**

An interesting study on the response of vegetation zones to climate change, using the aDGVM (adaptive Dynamic Global Vegetation Model).

**Seidl, R., P. M. Fernandes, and T. F. Fonseca, et al. 2011. Modelling natural disturbances in forest ecosystems: A review.** *Ecological Modelling* **222.4: 903–924.**

A review on forest models with a focus on stochastic disturbances.

## Metacommunity, Biogeography, and Macroecological Models

In the early 21st century, there has been increasing appreciation of the importance of stochasticity for the emergence of diversity, distribution, and co-occurrence patterns. This is not least due to the debate about niche vs. neutral processes reviewed in the separate *Oxford Bibliographies* article Niche Versus Neutral Models of Community Organization. As a result, stochastic simulations are frequently employed to study metacommunity, biogeographical, and macroecological dynamics. A comprehensive review of simulation models for the entire field is provided in Cabral, et al. 2017. Interesting studies include Pulliam 2000, which presents a simulation model to study the relationship between niche and distribution. Kearney and Porter 2009 presents a mechanistic niche model, and Chave, et al. 2002 presents a simulation model to study the diversity patterns emerging in neutral and non-neutral metacommunities. Further modeling examples can be found in the separate *Oxford Bibliographies* articles Metacommunity Dynamics and Assembly Models

**Cabral, J. S., L. Valente, and F. Hartig. 2017. Mechanistic simulation models in macroecology and biogeography: State-of-the-art and prospects.** *Ecography* **40:267–280.**

A comprehensive review of simulation models in macroecology and biogeography.

**Chave, J., H. C. Muller-Landau, and S. A. Levin. 2002. Comparing classical community models: Theoretical consequences for patterns of diversity.** *American Naturalist* **159.1: 1–23.**

Uses a simulation model for studying the diversity patterns emerging from neutral and non-neutral metacommunities. The simulations show how species area curves and rank abundance distributions change, depending on the assumed underlying ecological processes.

**Kearney, M., and W. Porter. 2009. Mechanistic niche modelling: Combining physiological and spatial data to predict species' ranges.** *Ecology Letters* **12.4: 334–350.**

The study introduces a framework for creating mechanistic niche models based on physiological processes.

**Pulliam, H. R. 2000. On the relationship between niche and distribution.** *Ecology Letters* **3.4: 349–361.**

Uses a simulation model to study the relationship between niche and distribution.

## Movement and Collective Behavior

Animal movement and collective behavior is another area where both stochasticity and discrete entities favor the application of simulation models. Collective behavior refers to the processes leading to swarming and flocking. An excellent example is Couzin, et al. 2005. Potts and Lewis 2014 concludes that simulation models are also frequently employed to study the formation of home ranges and territories. A further application is the estimation of connectivity, which is important for wildlife conservation. An example for this branch of models is Palmer, et al. 2011.

**Couzin, I. D., J. Krause, N. R. Franks, and S. A. Levin. 2005. Effective leadership and decision-making in animal groups on the move.** *Nature* **433.7025: 513–516.**

An influential study on the individual behavior and decision rules leading to the formation of animal groups that move collectively.

**Palmer, S. C., A. Coulon, and J. M. Travis. 2011. Introducing a 'stochastic movement simulator' for estimating habitat connectivity.** *Methods in Ecology and Evolution* **2.3: 258–268.**

An example of a simulation model to estimate habitat connectivity by simulating movement rules.

**Potts, J. R., and M. A. Lewis. 2014. How do animal territories form and change? Lessons from 20 years of mechanistic modelling.** *Proceedings of the Royal Society of London B: Biological Sciences* **281.1784: 20140231.**

A review of models that describe the formation of animal territories.

## Evolutionary and Macroevolutionary Models

Finally, simulation models have been highly influential in developing evolutionary theory. Examples are Nowak and May 1992, which presents spatial simulations on the evolution of cooperation, and Szabó and Fath 2007, which studies the same problem with more complex interaction structures. Simulation models are also commonly applied in macroevolution. An example for the latter is Colwell and Rangel 2010. Further examples are reviewed in Cabral, et al. 2017.

**Cabral, J. S., L. Valente, and F. Hartig. 2017. Mechanistic simulation models in macroecology and biogeography: State-of-the-art and prospects.** *Ecography* **40:267–280.**

This review of macroecological simulation models also covers macroevolutionary simulations with an ecological component.

**Colwell, R. K., and T. F. Rangel. 2010. A stochastic, evolutionary model for range shifts and richness on tropical elevational gradients under Quaternary glacial cycles.** *Philosophical Transactions of the Royal Society of London B: Biological Sciences* **365.1558: 3695–3707.**

This study employs an evolutionary simulation to understand macroevolutionary patterns.

**Nowak, M. A., and R. M. May. 1992. Evolutionary games and spatial chaos.** *Nature* **359.6398: 826–829.**

This classic paper is concerned with understanding the evolution of cooperation through spatial simulations.

**Szabó, G., and G. Fath. 2007. Evolutionary games on graphs.** *Physics Reports* **446.4: 97–216.**

This paper reviews the extension of spatial evolutionary simulations to more complex interaction structures that can be described by graphs.

---

# Methodology

To work with simulation models, ecologists need to implement them in a computer program and test whether this implementation behaves as expected. They may then want to analyze model reactions to parameter changes, calibrate model parameters, test whether the model fits to observed data, and make forecasts. This section summarizes key references of the methods applied for these purposes.

## Programming Languages and Software

To implement a computer model, one needs a programming language. For complex models, where flexibility and computational time is key, one usually employs general-purpose languages such as Fortran, C/C++, and Java. For prototyping models, smaller simulations, or for the analysis of simulation results, the scientific programming languages R, Matlab, and Python are common choices. A recent competitor for these languages is Julia, which combines a clean and easy syntax with excellent performance for stochastic simulations. Finally, there many programs and libraries that provide dedicated environments for particular classes of simulations. Two popular examples are NetLogo for individual-based/agent-based models, and Vensim for system-dynamics models.

**C/C++.**

C and its object-oriented counterpart C++ are the standard general-purpose languages for performance-critical programs. Moreover, the interpreters of many higher-level interpreted languages, such as R and Python, are coded in C.

**Fortran.**

Before the advent of C/C++, Fortran used to be the traditional language of scientific computing. It is still used in many older models, as Fortran offers excellent scientific libraries and a computation speed on par with C/C++. However, due to a more limited set of functionalities compared to C/C++, new projects will typically use the latter.

**Java.**

Java is generally slower than C/C++, but has a stronger and more consistent object-orientation. As such, it has become a common choice for simulation models that can use these advantages, in particular individual-based models.

**Julia.**

Julia has emerged as a competitor to R, Matlab, and Python. One of the main advantages of Julia is that it is optimized for speed when performing rule-based simulations. Thus, it seems an interesting choice for smaller simulation models that would otherwise be implemented in C.

**Matlab.**

Matlab, a commercial program, is a popular choice for scientific computing in the physical sciences. It is also widely used in machine-learning, complexity sciences, and signal processing.

**NetLogo.**

NetLogo is a simulation environment mainly aimed at agent-based/individual-based simulations. It supports simulations on grids, continuous space, and networks, and comes with a large library of example models that make it an excellent tool for teaching.

**Python.**

Python, another high-level scripting language, is currently one of the most popular languages for bioinformatics and systems biology.

**R.**

R is the most popular language for statistical computing. Although it would be unlikely that larger simulation models are coded in R, it is used both for the analysis of simulations and for coupling to models that are written in lower-level languages.

**Vensim.**

Vensim is commercial simulation software for system dynamics modeling. There are many similar commercial and free software solutions for this purpose, another popular one being STELLA.

## Implementation and Verification

After the choice of the programming language/simulation environment, the model needs to be implemented and verified, the latter meaning to confirm that the technical implementation of the model is correct. As scientists are usually not trained in programming, it is common that technical errors slow down the implementation process. Wilson 2006 gives some excellent hints for techniques from computer science that should be more widely adopted by scientists working in scientific computing. A review for best practice in scientific computing is provided in Wilson, et al. 2014.

**Wilson, G. V. 2006. Where's the real bottleneck in scientific computing?** *American Scientist* **94.1: 5.**

A readable opinion piece arguing for better educating scientists in the tools and best practices of computer science and programming.

**Wilson, G., D. A. Aruliah, C. T. Brown, et al. 2014. Best practices for scientific computing.** *PLoS Biology* **12.1: e1001745.**

A review on best practices for scientific computing.

## Sensitivity Analysis

After the implementation and verification of a model, it is common to perform a sensitivity analysis. A sensitivity analysis explores how a model reacts to changes in inputs or parameters. A recent overview is provided in Cariboni, et al. 2007. A comprehensive textbook is Saltelli, et al. 2000.

**Cariboni, J., D. Gatelli, R. Liska, and A. Saltelli. 2007. The role of sensitivity analysis in ecological modelling.** *Ecological Modelling* **203.1: 167–182.**

Introducing the importance and practice of sensitivity analysis for analyzing the behavior of ecological simulation models.

**Saltelli, A., K. Chan, and E. M. Scott, eds. 2000.** *Sensitivity analysis***. New York: Wiley.**

A comprehensive reference on local and global methods for sensitivity analysis.

## Calibration and Validation

Typically, not all parameters of a model can be fixed a priori. In this case, models need to be calibrated, meaning that parameters are adjusted such that the model outputs fit to observed data. For a deterministic model, if one is willing to assume independent normal errors, the problem can be treated as a complicated nonlinear regression with standard frequentist or Bayesian methods (e.g., van Oijen, et al. 2005). However, in practice, these assumptions are often either not met, or do not correspond to what modelers view as essential for the performance of a model. In a highly influential paper, Beven and Freer 2001 popularized the GLUE methodology, which essentially suggests summarizing the desired performance criteria into an objective function with no clear statistical interpretation but nevertheless use this objective in the place of the posterior in a Bayesian analysis. Another suggestion for model calibration is pattern-oriented modeling (POM, see Grimm, et al. 2005), which suggests that model predictions need to satisfy a number of user-defined performance criteria. For stochastic models, more rigorous statistical methods, such as Approximate Bayesian Computation and Synthetic Likelihood, have become popular. A review of these approaches is provided in Hartig, et al. 2011.

**Beven, K., and J. Freer. 2001. Equifinality, data assimilation, and uncertainty estimation in mechanistic modelling of complex environmental systems using the GLUE methodology.** *Journal of Hydrology* **249.1: 11–29.**

A highly influential paper that popularized the GLUE methodology, which attempts to define an arbitrary (but sensible) objective function and then essentially use Bayesian methods, in particular MCMC sampling, to analyze the response surface of this objective function.

**Grimm, V., E. Revilla, and U. Berger, et al. 2005. Pattern-oriented modeling of agent-based complex systems: Lessons from ecology.** *Science* **310.5750: 987–991.**

A key reference for the framework of pattern-oriented modeling, which proposes that good models or model parameters are characterized by the fact that model outputs respect a number of empirical patterns that have before been defined by the modeler.

**Hartig, F., J. M. Calabrese, B. Reineking, T. Wiegand, and A. Huth. 2011. Statistical inference for stochastic simulation models—theory and application.** *Ecology Letters* **14.8: 816–827.**

A review on methods to generate approximate likelihood functions for stochastic models, including Approximate Bayesian Computation (ABC) and synthetic likelihood approaches.

**Van Oijen, M., J. Rougier, and R. Smith. 2005. Bayesian calibration of process-based forest models: Bridging the gap between models and data.** *Tree Physiology* **25.7: 915–927.**

The study demonstrates how simulation models can essentially be used as complex regressions and thus be calibrated with standard (Bayesian) statistical methods.

## Forecasting and Uncertainty Analysis

A final step in the modeling process is making forecasts. A key issue in forecasting is to correctly forward uncertainty and stochasticity in inputs, processes, and parameters to the model outputs. This process, known as uncertainty analysis, is reviewed in Saltelli, et al. 2000. Clark, et al. 2003 is an example demonstrating that process stochasticity can have crucial influences on model forecasting uncertainty. Petchey, et al. 2015 discusses the question of forecasting uncertainty for ecology more generally. Dietze 2017 is a textbook concentrating entirely on forecasting with ecological models.

**Clark, J. S., M. Lewis, J. S. McLachlan, and J. HilleRisLambers. 2003. Estimating population spread: What can we forecast and how well?** *Ecology* **84.8: 1979–1988.**

A demonstration of the extent of forecasting uncertainty that can be created by stochasticity and parametric uncertainty.